\documentclass{article}
\usepackage{spconf,amsmath,amssymb,graphicx,nicefrac}

\newcommand{\GL}[1][3]{\text{GL}^+(#1)}

\DeclareMathOperator{\myexp}{exp}
\DeclareMathOperator{\mean}{mean}

\title{Towards Shape-based Knee Osteoarthritis Classification using\\Graph Convolutional Networks}
\name{Christoph von Tycowicz}
\address{Zuse Institute Berlin, Berlin, Germany}
\begin{document}
%
\maketitle
\begin{abstract}

We present a transductive learning approach for morphometric osteophyte grading based on geometric deep learning.
We formulate the grading task as semi-supervised node classification problem on a graph embedded in shape space.
To account for the high-dimensionality and non-Euclidean structure of shape space we employ a
combination of an intrinsic dimension reduction together with a graph convolutional neural network.
We demonstrate the performance of our derived classifier in comparisons to an alternative extrinsic approach.

\end{abstract}

%
%
\section{Introduction}
\label{sec:intro}

Osteoarthritis (OA) is a joint disease associated with defective integrity of articular cartilage and related changes in the underlying bone and at the joint margins. With over 250 million people affected world wide, OA ranks globally among the 50 most common sequelae of diseases and injuries~\cite{vos2012years}. Of the global disease burden, over 80\% involve the tibiofemoral joint of the knee.
In clinical practice plain radiography remains a mainstay for the diagnosis of OA with the Kellgren-Lawrence (KL) grading system~\cite{kellgren1957KLscore} being the de-facto standard classification scheme.
However, radiographic approaches suffer from several disadvantages:
First, although the evaluation of anatomic changes is an inherently three-dimensional problem, the radiographs only provide two-dimensional projections. 
Second, cartilage degeneration is only visible indirectly in terms of joint-space narrowing and bony changes requiring highly experienced practitioners.
The overall socio-economic burden associated with OA provides a strong impetus to develop novel computer-aided diagnostics 
that provide objective tools to support clinical decision-making.\newline 
%
%
Automatic knee OA diagnosis has a long history starting as early as 1989~\cite{dacree1989automatic} with the majority of works focusing on the analysis of plain radiographic images~\cite{shamir2010assessment,woloszynski2012dissimilarity,oka2008fully}.
Accessibility of large-scale clinical studies such as Osteoarthritis Initiative (OAI) further spurred the developments in particular using approaches from deep learning~\cite{antony2016quantifying,antony2017automatic} already reaching human-level diagnostic performance~\cite{tiulpin2018automatic}.\newline
However, due to the limitations of X-ray imaging, radio\-graphy-based OA assessment is insensitive when attempting to detect changes in early OA~\cite{tiulpin2018automatic} and potentially suffers from significant variability due to varying imaging settings and data acquisition set-ups.
Shape-based approaches \cite{bredbenner2010statistical,barr2012predictingOA,neogi2013predictOA}, on the other hand, hold the promise of increased robustness toward such variability and, hence, are better transferable across different datasets.
As shapes are elements of nonlinear spaces that carry a rich geometric structure, linear approaches are limited in capturing the complex structural variability evident in a population.
Contrary, being faithful to this nonlinear structure has been shown to provide highly consistent results in particular w.r.t. to OA classification~\cite{vonTycowicz2018DCM,Ambellan2019FCM,Ambellan2019GLplus}.\newline
In recent years, machine learning and in particular deep learning has been proven to be very successful for a variety of data analysis tasks in medical image computing and beyond.
However, these techniques have been most successful on data featuring a Euclidean or regular grid-like structure such as, e.g., radiographic images.
Due to the non-Euclidean nature of shape space, these constructions are not easily transferred as there are no such familiar properties like a global system of coordinates or shift-invariance.
Approaches that generalize deep neural models to non-Euclidean domains in order to leverage the intrinsic structure belong to the field of geometric deep learning and we refer to~\cite{bronstein2017geometric} for an overview.\newline
Beside their application domain, learning procedures can be classified into inductive and transductive approaches.
While inductive learning tries to infer a general model from labeled examples in order to predict labels of unseen ones, transductive approaches learn labels simultaneously on training and test data and, thus, can utilize training patterns directly while deciding for a test pattern.
Transductive learning therefore avoids solving a more general problem as an intermediate step and thus faces a simpler problem as compared to inductive learning.\newline
In this work, we derive a novel transductive learning approach for automatic grading of osteophytes from morphometric knee bone data.
Our approach utilises deep neural networks for anatomical shape data combining concepts from computational anatomy and geometric deep learning.
To the best of our knowledge, we are the first to present a geometric deep learning classifier in this field, paving the way for novel approaches to computer-aided diagnosis of knee osteoarthritis.

\begin{figure*}[tb]
    \center
    \includegraphics[width=0.99\linewidth]{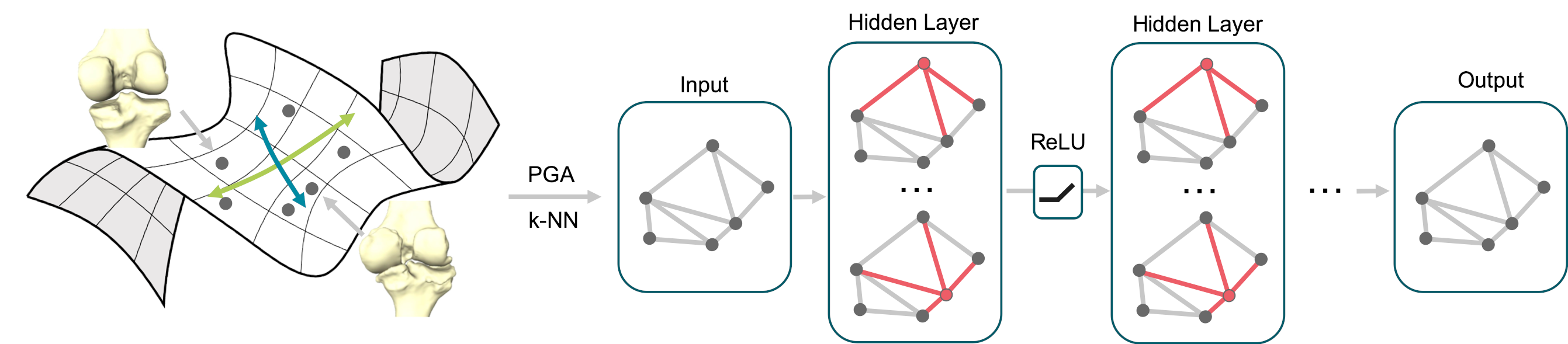}
    \caption{Schematic depiction of semi-supervised osteophyte classification procedure applying a graph convolutional neural network on a sub-manifold embedded in shape space.}
    \label{fig:pipeline}
\end{figure*}

%
%
\section{Method}
\label{sec:method}

This section introduces the methodology proposed for the automatic grading of osteophytes from shape data.
In particular, we consider shapes of knee bones (distal femur and proximal tibia) as elements of shape space, i.e.\ a high-dimensional and curved manifold.
Although such biological shapes feature high natural variability, instances of a certain object class will lie in a significantly smaller sub-space.
We obtain a structural description of this underlying sub-manifold by constructing a graph that encodes the heterogeneous pairwise relationships in the shape data.
To this end, we construct an undirected k-nearest neighbors graph $\mathcal{G}$ based on geodesic distances $\{d_{ij}\}$ in shape space.
Let $W$ be the corresponding (weighted) adjacency matrix with entries $W_{ij} = \myexp(-d_{ij}^2/\varepsilon)$ and kernel bandwidth $\varepsilon = \mean(\{d^2_{ij}\})^{3/4}$.
In the remainder of this section, we first provide details on the employed shape model and graph signal processing before we derive our approach to transductive OA classification.
An overview of the entire pipeline is shown in Fig.~\ref{fig:pipeline}

\subsection{Shape Space}
\label{sec:shapespace}

In medical image analysis shape usually refers to the boundary of anatomical objects belonging to a particular class of tissues so that they can be represented as deformations of a common template surface $\mathcal{M}$.
In this context, shape representations based on differential coordinates~\cite{vonTycowicz2018DCM} have been proven to provide highly discriminatory descriptions of shape.
The main idea is to encode a deformation $\phi: \mathcal{M} \rightarrow \mathbb{R}^3$ and thereby a shape in terms of its derivative $\nabla\phi$ called deformation gradient.
At any point $x \in \mathcal{M}$, the gradient $\nabla\phi|_x \in \GL$ can be factored uniquely into a rotational $R$ and stretch $U$ component using the well-known polar decomposition $\nabla\phi|_x = RU$.
This decomposition gives rise to a physically-motivated distance that effectively captures local changes in shape.
Remarkably, despite its nonlinear structure this model exhibits closed-form expressions allowing for simple and efficient computations.


\subsection{Graph Convolutional Filters}

A central ingredient in obtaining convolutional architectures on graph-structured data is to derive signal filtering operators that generalize convolutions.
In our derivation, we opt for a spectral approach over a spatial one as the latter require local charting which is not readily available on non-Euclidean domains.
Motivated by the convolution theorem, spectral generalizations model graph convolutions as linear operators that commute with the (normalized) graph Laplacian $\Delta = I - D^{-1/2}WD^{-1/2}$, where $D$ is the diagonal degree matrix with $D_{ii} = \sum_j W_{ij}$.
However, such filters are not naturally localized and computationally costly as they depend on the spectral transform.
These issues can be overcome by using a truncated expansion of the spectral filter in terms of Chebyshev polynomials~\cite{defferrard2016convolutional} up to $K^\text{th}$-order.
In particular, such a parametrization results in $K$-localized filters, i.e.\ the output depends only on the $K$-hop neighborhood at each node.
Furthermore, this approach is efficient to evaluate and learn as it allows for a spectrum-free formulation.
In our approach, we employ a first-order parametrization yielding graph filter $g_\theta$ with learnable parameters $\theta$ of the form
\begin{align}\label{eq:convolution} \nonumber
    g_\theta(\Delta) = \theta_0 I + \theta_1 \tilde{\Delta},
\end{align}
where $\tilde{\Delta} = 2\Delta/\lambda_\text{max}-I$ and $\lambda_\text{max}$ denote the scaled graph Laplacian and the largenst eigenvalue of $\Delta$, respectively.



\subsection{Semi-supervised Classification}

Based on the graph representation $\mathcal{G}$,
we can cast the problem of transductive osteophyte grading as node classification task, where labels are only available for a subset of the nodes.
Employing graph convolutional filters, we are able to avoid explicit graph-based regularization by encoding the graph structure directly via a graph convolutional neural network model and training it on a supervised loss $\mathcal{L}$ that takes only labeled nodes into account.
As the model is conditioned on the adjacency matrix of the graph, the model is able to learn representations of both labeled and unlabeled nodes as gradient information from $\mathcal{L}$ can be distributed throughout the model.

In particular, we construct a multi-layer, feed-forward graph convolutional network with possibly several hidden layers each followed by a rectified linear unit (ReLU).
The number of final feature channels is constrained to the desired number of classes.
The loss energy is computed by a node-wise soft-max over the final feature map combined with the cross entropy loss function evaluated only on labeled nodes.
We train the network using the RMSprop otimizer performing batch descent with the full dataset for every training iteration.
To introduce stochasticity during training we add dropout (0.25 probability) to each layer.

As shapes expose many degrees of freedom, employing their high-dimensional representation as input features would yield a high-capacity model destined to overfitting.
We circumvent this issue by performing a Riemannian dimension reduction to extract a low-dimensional descriptor that encodes the intrinsic geometric structure.
To this end, we perform \textit{principal geodesic analysis}~\cite{Fletcher2004PGA} to estimate a geodesic sub-manifold that best captures the variability in the input shapes.

%
%
\section{Experiments and Results}
\label{sec:experiments}

\subsection{Data description}

Our shape data comprises distal femora and proximal tibiae for 201 (67 per class) randomly selected subjects from the OsteoArthritis Initiative (a longitudinal, prospective study of knee OA) for which the segmentations are publicly available\footnote{https://doi.org/10.12752/4.ATEZ.1.0}~\cite{ambellan2019automated}.
The subjects are independently graded with 0 to 2 indicating none, minute, and definite presence of osteophytes, respectively.

\subsection{Experimental Setup}

For experimental evaluation we employed a model as described in Sec.~\ref{sec:method} with 3 layers and 64 feature channels (determined via hyper-parameter optimization).
To control for overfitting, the data was split into training, validation, and test sets with a ratio of $\nicefrac{2}{3}$, $\nicefrac{1}{6}$, and $\nicefrac{1}{6}$, respectively allowing for 3-fold cross-validation. We trained the model for 500 epochs choosing the best configuration according to validation accuracy.
To account for randomness, we performed cross-validation on 100 permutations and report average testing accuracies.
We further compare our intrinsic graph convolutional network to an extrinsic version that employs a flat Euclidean metric on shape space (with 2 layers and 96 feature channels according to hyper-parameter optimization). 

\subsection{Results}

The average test set multi-class accuracy achieved by our intrinsic model was 64.64\%.
In contrast, comparing this result to the Euclidean approach, which averaged at 58.62\%, this is a substantial improvement in classification accuracy.

%
%
\begin{table}[h!]
\centering
\caption{Confusion matrix between our GCN's predictions and the ground truth osteophyte classes.}
\begin{tabular}{c|ccc}
prediction \textbackslash~actual class  & 0   & 1   & 2          \\ \hline
0                                & 64.6\% & 30.2\% & 13.2\%      \\
1                                & 28.1\% & 54.4\% & 15.1\%      \\
2                                & ~~7.3\%  & 15.4\% & 71.7\%      \\
\end{tabular}                               
\label{tab:results_confusion_matrix}
\end{table}
%
%

A more differentiated overview of the accuracy of our intrinsic model is given by the confusion matrix presented in Table~\ref{tab:results_confusion_matrix}.
Consitently to the ordinal nature of osteophyte grades the off-diagonal entries in the confusion matrix decay with the distance to the diagonal.
We further assess the agreement of our model with the expert annotations from the OAI dataset using Cohen's quadratically weighted kappa coefficient.
In particular, the kappa value on the test set was 0.58 indicating a moderate agreement.

%
%
\section{Conclusion}
\label{sec:conclusion}

We presented a novel shape-based deep learning approach to automatic knee osteophyte grading.
A key feature of our method is that it is formulated intrinsically, i.e.\ it is coordinate-free and does not suffer from linearization errors.
Furthermore, we propose a transductive approach that is conceptually simpler to classical inductive methods promising improved classification accuracies.
As our graph neural network is fully convolutional, it can be directly applied to unseen data.
In future work, we will therefore investigate the predictive power of such an out-of-sample approach in contrast to a full transductive re-training.
Another interesting direction will be to employ a shape-based representation of the joint gap. 
This would allow to also predict narrowing which would pave the way towards a morphometric, computer-aided diagnosis of knee OA.

\bibliographystyle{IEEEbib}
\bibliography{article}

\end{document}